\documentclass
[%
	draftclsnofoot,  % [draft, draftcls, draftclsnofoot, >final] cls=yes figures, nofoot=no draft footnote
	peerreviewca,  % [conference, >journal, technote, peerreview, peerreviewca] ca=for conference mode
]%
{IEEEtran}%

\IEEEoverridecommandlockouts
\usepackage{cite}
\usepackage{amsmath,amssymb,amsfonts}
\usepackage{algorithmic}
\usepackage{graphicx}
\usepackage{textcomp}
\usepackage{xcolor}

\graphicspath{{figures/}{images/}{fig/}}  % search paths for graphics
\DeclareGraphicsExtensions{.pdf,.jpeg,.png}

\ifCLASSOPTIONcompsoc
	\usepackage
	[%
		caption=false,
		font=normalsize,
		labelfont=sf,
		textfont=sf,
	]{subfig}
\else
	\usepackage
	[%
		caption=false,
		font=footnotesize,
	]{subfig}
\fi

\usepackage{tikz}
\usetikzlibrary
{%
	patterns,
	positioning,
	decorations.pathreplacing,
}%

\usepackage[hidelinks]{hyperref}	% Provides clickable links in the PDF-document for \ref
\usepackage{siunitx}				% Units with correct spacing. \SI{12}{\dB}, \SI{4}{\mega\hertz}, \ang{90}, \num{1e-3}, \num{10000}
\usepackage{xspace}					% Intelligent spacing after text macros

\usepackage{enumitem}				% resume enumeration across different environments

\input{99_custommacros.sty}

\newcommand{\timesteptext}{time~step~\( \timeindex \)\xspace}
\newcommand{\usertext}{user~\( \userindex \)\xspace}

\newcommand{\internalemergencyvehicle}{EV}
\newcommand{\emergencyvehicletext}{\textsc{\small \internalemergencyvehicle}\xspace}
\newcommand{\emergencyvehicleformula}{\text{\textsc{\internalemergencyvehicle}}}

\newcommand{\maximumthroughput}{\textsc{\small MT}\xspace}
\newcommand{\delaysensitive}{\textsc{\small DS}\xspace}
\newcommand{\maxminfair}{\textsc{\small MMF}\xspace}

\newcommand{\channelqualityuser}[1][\userindex]{\internalchannelquality_{#1}[\timeindex]}
\newcommand{\channelqualityrayleighuser}{\tilde{\internalchannelquality}_{\userindex}[\timeindex]}
\newcommand{\variancechannelquality}{\variance_{\internalchannelquality}}

\newcommand{\userindex}							{n}
\newcommand{\numusers}							{N}
\newcommand{\jobindex}							{j}
\newcommand{\numjobs}							{J}
\newcommand{\internalresourceblock}				{u}
\newcommand{\numresourcesavailable}				{U}
\newcommand{\internaldistanceuser}				{d}
\newcommand{\chancejobs}						{c_{\jobindex}}
\newcommand{\internaljobsset}{\mathbb{\numjobs}}
\newcommand{\timeindexalt}						{\tau}
\newcommand{\userindexalt}{\tilde{\userindex}}
\newcommand{\resourceblockrequestedjob}[1][\jobindex]{\internalresourceblock_{{#1}, \, \text{req}}[\timeindex]}
\newcommand{\resourceblockrequestedjobalt}[1][\jobindex]{\internalresourceblock_{{#1}, \, \text{req}}[\timeindexalt]}
\newcommand{\resourceblockscheduleduser}{\internalresourceblock_{\userindex, \, \text{sx}}[\timeindex]}
\newcommand{\resourceblockmaxuser}{\internalresourceblock_{\userindex, \, \text{max}}}
\newcommand{\jobsset}{\internaljobsset[\timeindex]}
\newcommand{\jobsuserset}[1][\timeindex]{\internaljobsset_{\userindex}[#1]}
\newcommand{\jobsnewset}[1][\timeindex]{\internaljobsset_{n,\,\text{new}}[#1]}
\newcommand{\jobsfailedset}{\internaljobsset_{\text{fail}}[\timeindex]}
\newcommand{\distanceuser}{\internaldistanceuser_{\userindex}[\timeindex]}

\newcommand{\internaltimetotimeout}				{v}
\newcommand{\internallowesttimetotimeout}		{l}
\newcommand{\internaltotaltimeouts}				{m}
\newcommand{\internalpacketrate}				{k}
\newcommand{\timetotimeoutjob}{\internaltimetotimeout_{\jobindex}[\timeindex]}
\newcommand{\timetotimeoutinitialuser}{\internaltimetotimeout_{\userindex,\,\text{init}}}
\newcommand{\lowesttimetotimeoutuser}[1][\userindex]{\internallowesttimetotimeout_{#1}[\timeindex]}
\newcommand{\totaltimeouts}[1][\userindex]{\internaltotaltimeouts_{#1}[\timeindex]}
\newcommand{\packetrateuser}[1][\userindex]{\internalpacketrate_{#1}[\timeindex]}

\newcommand{\reward}[1][]{\internalreward_{#1}[\timeindex]}
\newcommand{\rewardtimeout}{\internalreward_{L}[\timeindex]}
\newcommand{\rewardtimeoutev}{\internalreward_{L, \emergencyvehicleformula}[\timeindex]}
\newcommand{\rewardcapacity}{\internalreward_{C}[\timeindex]}
\newcommand{\rewardpacketrate}{\internalreward_{P}[\timeindex]}
\newcommand{\rewardtimeoutuser}[1][\timeindex]{\internalreward_{L,\,\userindex}[#1]}
\newcommand{\rewardestimate}[1][]{\hat{\internalreward}_{#1}[\timeindex]}
\newcommand{\weightrewardtimeout}{\weight_{L}}
\newcommand{\weightrewardtimeoutev}{\weight_{L, \emergencyvehicleformula}}
\newcommand{\weightrewardcapacity}{\weight_{C}}
\newcommand{\weightrewardpacketrate}{\weight_{P}}

\newcommand{\statescauser}[1][]{\internalstate_{\userindex, {#1}}[\timeindex]}

\newcommand{\statevec}[1][]{\mathbf{\internalstate}_{#1}[\timeindex]}
\newcommand{\actionvec}[1][]{\mathbf{\internalaction}_{#1}[\timeindex]}
\newcommand{\actionvecnoisy}[1][]{\hat{\mathbf{\internalaction}}_{#1}[\timeindex]}
\newcommand{\parametersactor}{\parameters_{\actornetworkvec}[\timeindex]}
\newcommand{\parameterscritic}{\parameters_{\longrewardestimsca}[\timeindex]}

\newcommand{\internalpercentage}				{A}
\newcommand{\percentageuser}[1][\userindex]{\internalpercentage_{#1}[\timeindex]}
\newcommand{\distortionuser}{\tilde{\internalpercentage}_{\userindex}}
\newcommand{\explorationnoisemultiplier}{\variance_{\text{e}}}

\newcommand{\losscritic}{\internalloss_{\longrewardestimsca}}
\newcommand{\lossactor}{\internalloss_{\actornetworkvec}}

\newcommand{\channelpriority}{\bar{\internalchannelquality}_{\userindex}[\timeindex]}
\newcommand{\timeoutpriority}{\bar{\internaltotaltimeouts}_{\userindex}[\timeindex]}
\definecolor{ccolor1}{RGB}{208,27,136}
\definecolor{ccolor2}{RGB}{37,71,150}
\definecolor{ccolor3}{RGB}{48,123,59}
\definecolor{ccolor4}{RGB}{202,160,35}
\begin{document}

\title
{%
	Learning Resource Scheduling with High Priority Users using Deep Deterministic Policy Gradients
	\thanks
	{%
		This work was partly funded by the German Ministry of Education and Research (BMBF) under grant 16KIS1028 (MOMENTUM). 
		This work was accepted for presentation at IEEE ICC 2022.
	}%
}%

\author{%
	\IEEEauthorblockN{%
		Steffen~Gracla%
		% \,\orcidlink{0000-0003-3315-9280}%
		,
		Edgar~Beck%
		% \,\orcidlink{0000-0003-2213-9727}%
		,
		Carsten~Bockelmann%
		% \,\orcidlink{0000-0002-8501-7324}%
		\ and
		Armin~Dekorsy%
		% \,\orcidlink{0000-0002-5790-1470}%
	}%
	\IEEEauthorblockA{%
		Dept. of Communications Engineering, University of Bremen, Bremen, Germany\\
		{Email: \{gracla, beck, bockelmann, dekorsy\}@ant.uni-bremen.de}
	}%
}%

\maketitle

\begin{abstract}
		Advances in mobile communication capabilities open the door for closer integration of pre-hospital and in-hospital care processes.
		For example, medical specialists can be enabled to guide on-site paramedics and can, in turn, be supplied with live vitals or visuals.
		Consolidating such performance-critical applications with the highly complex workings of mobile communications requires solutions both reliable and efficient, yet easy to integrate with existing systems.
		This paper explores the application of Deep Deterministic Policy Gradient~(\ddpg) methods for learning a communications resource scheduling algorithm with special regards to priority users.
		Unlike the popular Deep-Q-Network methods, the \ddpg is able to produce continuous-valued output.
		With light post-processing, the resulting scheduler is able to achieve high performance on a flexible sum-utility goal.
\end{abstract}


%% EXAMPLE TABLE
%\begin{table}[!t]
%	\renewcommand{\arraystretch}{1.3}
%	\caption{A Simple Example Table}
%	\label{tab:table_example}
%	\centering
%	\begin{tabular}{c||c}
%		\hline
%		\bfseries First & \bfseries Next\\
%		\hline\hline
%		1.0 & 2.0\\
%		\hline
%	\end{tabular}
%\end{table}

%% EXAMPLE FIGURE
%\begin{figure}[!t]
%	\centering
%	\includegraphics[width=2.5in]{myfigure}
%	\input{figures/myfigure.tex}
%	\caption{text}
%	\label{fig:figure_example}
%\end{figure}

\section{Introduction}\label{sec:intro}
The increased capabilities of mobile data transfer have opened opportunities to more tightly connect the chain of emergency care, \eg transmitting video, vitals or specialist input directly at an emergency site.
When time is of the essence, \cite{zanatta2016pre} show that a patients recovery chances may be improved significantly.
Pilot projects such as~\cite{buscher2014telemedical} have taken first steps in the direction of harnessing this potential of communications technology, but highly integrated solutions are only just emerging.
This is in part due to the substantial performance demands that medical applications require in, \eg throughput, latency, and mean failure time.
These demands are outliers even in light of the 5G NR specifications of 3GPP~\cite{3gpp.22.826}.
Serving medical use cases therefore requires new, tailored solutions that are able to deal with complex optimization problems, balancing heterogeneous performance and reliability demands.

Deep Learning~(\deeplearning) methods have shown promising results dealing with highly complex tasks in a variety of fields including computer vision~\cite{krizhevsky2012imagenet} and human speech recognition~\cite{dahl2010phone}.
In these applications a data-driven learning approach has shown strengths where classic, model-based algorithms are struggling; Where an explicit, adjustable model is not required or feasible, \deeplearning offers algorithms that are approximately optimal for given resources and expected input data range~\cite{sun2017learning, mnih2015human}.
However, data-driven approaches must be applied in ways that are mindful of their limitations. 
Typical gradient-based \deeplearning methods result in approximate, hard-to-interpret algorithms in which undesired behavior patterns are hard or impossible to adjust.
For this reason, the applicability of \deeplearning for applications with strict performance demands requires further research, shown by, \eg \cite{gu_knowledge-assisted_2021}.

We consider a scenario where an Emergency Vehicle~(\emergencyvehicletext) shares a communication medium with normal users.
Through the use of intelligent resource scheduling, we can influence service quality for all users and ensure the proper handling of \emergencyvehicletext communications, though the question of QoS-optimal resource allocation for multiple traffic classes remains an open topic in research~\cite{nasralla_hybrid_2020}.
Other works, such as~\cite{ye2019deep, al2020learn}, have applied Deep Q-Networks~(\dqn) to the task of intelligent resource allocation in communications with some success.
While \dqn are able to leverage the function approximation capabilities of deep neural networks, the basic \dqn method restricts itself, by design, to discrete decision spaces.
Some problems, such as allocating available resources by proportion, lend themselves more easily to continuous-valued formulations.
In this paper, we investigate a Deep Deterministic Policy Gradient~(\ddpg)~\cite{lillicrap2015continuous, huang2020deep} based approach to learning a flexible resource allocation algorithm, directly outputting the proportion of resources to be allocated to each user.
In this way, the optimization process can be melded with the given problem more freely and easily compared to discrete action spaces.
Using these Reinforcement Learning~(\reinforcementlearning) methods, we form a scheduling algorithm that is able to drastically improve priority user performance with a minimal drop in the overall utility.

%% EXAMPLE TABLE
%\begin{table}[!t]
%	\renewcommand{\arraystretch}{1.3}
%	\caption{A Simple Example Table}
%	\label{tab:table_example}
%	\centering
%	\begin{tabular}{c||c}
%		\hline
%		\bfseries First & \bfseries Next\\
%		\hline\hline
%		1.0 & 2.0\\
%		\hline
%	\end{tabular}
%\end{table}

%% EXAMPLE FIGURE
%\begin{figure}[!t]
%	\centering
%	\includegraphics[width=2.5in]{myfigure}
%	\input{figures/myfigure.tex}
%	\caption{text}
%	\label{fig:figure_example}
%\end{figure}

\section{Setup \& Notation}
This section introduces the simulation environment that schedulers will interact with, characterized by a resource grid, a job queue, and the communication link to users.
We then formulate the optimization objective that motivates our \deeplearning-approach.

\subsection{Simulation Environment}
\begin{figure}[t]
	\centering
	\begin{tikzpicture}[scale=1.0]
\tikzstyle{rb} = [rectangle, draw, minimum width=.3cm, minimum height=.3cm]
\tikzstyle{rbu1} = [rb, pattern=north west lines, pattern color=ccolor2]
\tikzstyle{rbu2} = [rb, pattern=crosshatch dots, pattern color=ccolor1]
\tikzstyle{rbu3} = [rb, pattern=grid, pattern color=ccolor4]

\newcommand{\xanchor}{-1.0}
\newcommand{\xstep}{0.4}
\newcommand{\ystep}{0.4}

\node (r11) at (\xanchor + 0 * \xstep, +2.0 - 0 * \ystep) [rbu1] {};
\node (r12) at (\xanchor + 0 * \xstep, +2.0 - 1 * \ystep) [rbu2] {};
\node (r13) at (\xanchor + 0 * \xstep, +2.0 - 2 * \ystep) [rbu2] {};
\node (r14) at (\xanchor + 0 * \xstep, +2.0 - 3 * \ystep) [rbu2] {};
\node (r15) at (\xanchor + 0 * \xstep, +2.0 - 4 * \ystep) [rbu2] {};

\node(r21) [rb] at (\xanchor + 1 * \xstep, +2.0 - 0 * \ystep) {};
\node(r22) [rb] at (\xanchor + 1 * \xstep, +2.0 - 1 * \ystep) {};
\node(r23) [rb] at (\xanchor + 1 * \xstep, +2.0 - 2 * \ystep) {};
\node(r24) [rb] at (\xanchor + 1 * \xstep, +2.0 - 3 * \ystep) {};
\node(r25) [rb] at (\xanchor + 1 * \xstep, +2.0 - 4 * \ystep) {};

\node(r31) [rb] at (\xanchor + 2 * \xstep, +2.0 - 0 * \ystep) {};
\node(r32) [rb] at (\xanchor + 2 * \xstep, +2.0 - 1 * \ystep) {};
\node(r33) [rb] at (\xanchor + 2 * \xstep, +2.0 - 2 * \ystep) {};
\node(r34) [rb] at (\xanchor + 2 * \xstep, +2.0 - 3 * \ystep) {};
\node(r35) [rb] at (\xanchor + 2 * \xstep, +2.0 - 4 * \ystep) {};

\node(r41) [rb] at (\xanchor + 3 * \xstep, +2.0 - 0 * \ystep) {};
\node(r42) [rb] at (\xanchor + 3 * \xstep, +2.0 - 1 * \ystep) {};
\node(r43) [rb] at (\xanchor + 3 * \xstep, +2.0 - 2 * \ystep) {};
\node(r44) [rb] at (\xanchor + 3 * \xstep, +2.0 - 3 * \ystep) {};
\node(r45) [rb] at (\xanchor + 3 * \xstep, +2.0 - 4 * \ystep) {};

\node(r51) [rb] at (\xanchor + 4 * \xstep, +2.0 - 0 * \ystep) {};
\node(r52) [rb] at (\xanchor + 4 * \xstep, +2.0 - 1 * \ystep) {};
\node(r53) [rb] at (\xanchor + 4 * \xstep, +2.0 - 2 * \ystep) {};
\node(r54) [rb] at (\xanchor + 4 * \xstep, +2.0 - 3 * \ystep) {};
\node(r55) [rb] at (\xanchor + 4 * \xstep, +2.0 - 4 * \ystep) {};

\node(job111) [rbu1] at (-3.0, +2.2) {};
\node(job112) [rbu1] at (-2.6, +2.2) {};
\node(job113) [rbu1] at (-2.2, +2.2) {};
\node(job114) [rbu1] at (-3.0, +1.8) {};
\node(job115) [rbu1] at (-2.6, +1.8) {};

\node(job211) [rbu2] at (-3.0, +1.3) {};
\node(job212) [rbu2] at (-2.6, +1.3) {};
\node(job213) [rbu2] at (-2.2, +1.3) {};

\node(job221) [rbu2] at (-4.6, +1.3) {};
\node(job222) [rbu2] at (-4.2, +1.3) {};
\node(job223) [rbu2] at (-3.8, +1.3) {};
\node(job224) [rbu2] at (-4.6, +0.9) {};
\node(job225) [rbu2] at (-4.2, +0.9) {};
\node(job226) [rbu2] at (-3.8, +0.9) {};

\node(job311) [rbu3] at (-3.0, +0.4) {};
\node(job312) [rbu3] at (-2.6, +0.4) {};

\draw[]
	(\xanchor + 5 * \xstep, 0.2)
	-- node
		[align=center, above, rotate=-90]
		{$\numresourcesavailable = \num{5}$}
	(\xanchor + 5 * \xstep, 2.2);
	
\draw [-{stealth}]
	(\xanchor - 0.2, 0.0)
	-- node
		[below, align=left]
		{Time $\timeindex$}
	(\xanchor - 0.2 + 5 * \xstep, 0.0);
	
\draw[-]
	(-4.8, 0.1)
	--
	(-4.8, -0.1)
	-- node
		[below, align=center]
		{Job Set \( \jobsset \)}
	(-2.0, -0.1)
	--
	(-2.0, 0.1);
	
\draw [-{stealth}]
	(-1.8, 1.2)
	--
	(\xanchor - 0.3, 1.2);

\node(user1) [rbu1] at (-7.0, +2.2) {};
\node(user2) [rbu2] at (-7.0, +1.7) {};
\node(user3) [rbu3] at (-7.0, +1.2) {};
\node(user1t) [] at (-6.1, +2.2) {User 1};
\node(user2t) [] at (-6.1, +1.7) {User 2};
\node(user3t) [] at (-6.1, +1.2) {User 3};

% annotations
\draw [-] (-4.5, 1.5) -- +(-0.1, 0.2) node[above] {\( \resourceblockrequestedjob=\num{6} \)};

\end{tikzpicture}
	\caption{At each \timesteptext, a scheduler is tasked with distributing a number of~\( \numresourcesavailable \) discrete resources among users. In a job queue, jobs~\( \jobindex \) assigned to a \usertext are requesting a number~\( \resourceblockrequestedjob \) of discrete resources to be completed. This example shows a system configuration with \( {\numresourcesavailable = \num{5}} \)~managed resource blocks and \( {\numusers = \num{3}} \)~users.}
	\label{fig:resourceblocks}
\end{figure}
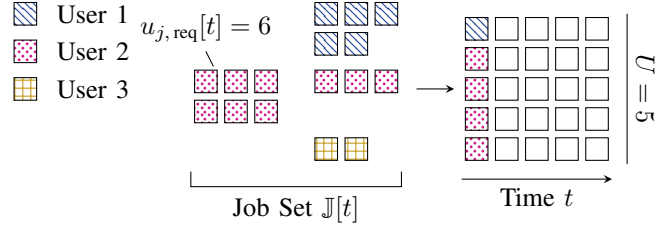

As depicted in \reffig{fig:resourceblocks}, we assume a scheduler at a base station that manages a limited number~\( \numresourcesavailable \) of discrete resource blocks, such as in an \ofdm-system.
In each discrete \timesteptext, the scheduler divides the available discrete resources~\( \numresourcesavailable \) among \( \numusers \)~connected users.
According to the fraction of resources that \usertext is granted, the resources are then filled with jobs~\( \jobindex \) from a job queue.
For performance metric calculation, we define the set~\( \jobsset \) of all jobs in queue at \timesteptext, and the set~\( \jobsuserset \) of the jobs assigned to \usertext.
At the beginning of each \timesteptext, every \usertext generates a new job~\( \jobindex \) with a probability~\( \chancejobs \), collected in sets~\( \jobsnewset \).
Jobs~\( \jobindex \) have two attributes: a remaining request size~\( \resourceblockrequestedjob \) in discrete resource blocks, and a time-to-timeout~\( \timetotimeoutjob \) in discrete time~steps~\( \timeindex \).
The jobs' initial values are set at generation, with the initial request size~\({
	\resourceblockrequestedjob \sim \mathbb{U}[\num{1}, \, \resourceblockmaxuser]
}\) drawn from a discrete uniform distribution, and the initial time-to-timeout~\({
	\timetotimeoutjob \gets \timetotimeoutinitialuser
}\) set to a fixed value~\( \timetotimeoutinitialuser \).
Both \( \resourceblockmaxuser \) and \( \timetotimeoutinitialuser \) are defined by a user-specific profile.

Once resources the have been divided in a \timesteptext, the appropriate amount of discrete blocks is deducted from the jobs' resource requests~\( \resourceblockrequestedjob \) and all remaining jobs' time-to-timeout~\( \timetotimeoutjob \) is decremented.
For record keeping, a lifetime count~\( \resourceblockscheduleduser \) of resources scheduled to a \usertext until \timesteptext is increased by the appropriate amount.
Should a job have remaining requests~\( {\resourceblockrequestedjob > \num{0}} \) when timing out, \ie~\( {\timetotimeoutjob = \num{0}} \), this job is removed from the queue and added to the set~\( \jobsfailedset \) of jobs that timed out during \timesteptext to be used in performance metric calculation.

At the beginning of a simulation episode, all user vehicles~\( \userindex \) are placed at a position drawn from a uniform random distribution centered on the base station position.
The user distance~\( \distanceuser \) then varies over time due to vehicle movement, which is simulated akin to the Manhattan-model of movement~\cite{aschenbruck2008survey}; At each \timesteptext, user vehicles move a unit-size step in a random direction, with a~\( \SI{98}{\percent} \) probability of selecting their previous direction and otherwise uniform probability of turning left, right, or stopping.
This results in a grid-like movement pattern.
The communication channel between the base station and the users~\( \userindex \) is assumed as a Rayleigh-fading channel with additional distance-proportional path loss.
Rayleigh-fading amplitudes~\({
	|\channelqualityrayleighuser| \sim \text{Rayleigh}(\variancechannelquality)
}\) are drawn from a Rayleigh-distribution with a scale~\( \variancechannelquality \), while the path loss~\( \pathloss \) is calculated according to the distance~\( \distanceuser \) between a \usertext and the base station at \timesteptext,
\begin{align}
	\pathloss
	=	\min
		\left(
			\num{1}, \,
			\left(\distanceuser\right)^{\num{-1}}
		\right)
	.
\end{align}
While typical path loss models assume exponents of~\(\num{-2}\) or less, we opted for an exponent of~\( \num{-1} \) to reduce spread, thereby decreasing the computation required in the learning task ahead.
Both path loss and Rayleigh-fading amplitude are combined into the power fading factor
\begin{align}
	\channelqualityuser
	=	| \channelqualityrayleighuser |
		\cdot
		\pathloss
	.
\end{align}

\subsection{Problem Statement}
\label{sec:problemstatement}
The scheduler will be tasked with learning an allocation strategy that balances performance in three global metrics: (a)~channel~capacity, (b)~timeouts, and (c)~packet rate.
These metrics will be defined subsequently.
We also introduce a priority class of users, the \emergencyvehicletext-type, that requires preferential treatment by the scheduler.

(a) We define the \emph{sum capacity}~\( \rewardcapacity \) at \timesteptext for a given user power fading~\( \channelqualityuser \), signal power~\( \signalpower \) and expected noise power~\( \noisepower \) as
\begin{align}
	\rewardcapacity
	=	\sum_{\userindex = 1}^{\numusers}
		\log
		\left(
			  \num{1}
			+ \channelqualityuser
			  \frac{\signalpower}{\noisepower}
		\right)
	=	\sum_{\userindex = 1}^{\numusers}
		\log
		\left(
			\num{1}
			+ \SNR_{\userindex}[\timeindex]
		\right)
	.
\end{align}
Signal power~\( \signalpower \) and expected noise power~\( \noisepower \) are assumed as fixed for all users.

(b) \emph{Sum timeouts}~\( \rewardtimeout \) are defined as the sum of resources among the set~\( \jobsfailedset \) of all jobs that have timed-out at the end of \timesteptext,
\begin{align}
	\rewardtimeout
	=	\sum_{\jobindex \in \jobsfailedset}
		\resourceblockrequestedjob
	.
\end{align}
We also define the metric~\( \rewardtimeoutev \) as the sum of resources lost from timeouts among \emergencyvehicletext-type users in \timesteptext.

(c) The \emph{sum packet rate}~\( \rewardpacketrate \) is the ratio of lifetime resources scheduled to a user~\( \resourceblockscheduleduser \), and the lifetime sum of resources requested by that \usertext.
It is calculated as
\begin{align}
	\label{eq:packetrate}
	\rewardpacketrate
	=	\sum_{\userindex = 1}^{\numusers}
		\frac
		{\resourceblockscheduleduser}
		{
			\sum_{\timeindexalt = 1}^{\timeindex}
			\sum_{\jobindex \in \jobsnewset[\timeindexalt]}
			\resourceblockrequestedjobalt
		}
	=	\sum_{\userindex = 1}^{\numusers}
		\packetrateuser
	.
\end{align}

To judge a scheduler's overall performance, the aforementioned metrics are combined into a sum utility metric
\begin{align}
\label{eq:reward}
	\reward
	=	  \weightrewardcapacity \rewardcapacity
		- \weightrewardtimeout \rewardtimeout
		- \weightrewardtimeoutev \rewardtimeoutev
		+ \weightrewardpacketrate \rewardpacketrate
	,
\end{align}
with individual tunable weights~\( \weightrewardcapacity, \weightrewardtimeout, \weightrewardtimeoutev, \weightrewardpacketrate \).
The weights' purpose is two-fold: Primarily, they balance the submetrics against each other, as the individual submetrics are not normalized in regard to their expected magnitude.
Secondarily, they allow us to communicate a preference or importance of the individual submetrics to the overall optimization process, such as in the case of \emergencyvehicletext-timeouts~\( \rewardtimeoutev \).

%% EXAMPLE TABLE
%\begin{table}[!t]
%	\renewcommand{\arraystretch}{1.3}
%	\caption{A Simple Example Table}
%	\label{tab:table_example}
%	\centering
%	\begin{tabular}{c||c}
%		\hline
%		\bfseries First & \bfseries Next\\
%		\hline\hline
%		1.0 & 2.0\\
%		\hline
%	\end{tabular}
%\end{table}

%% EXAMPLE FIGURE
%\begin{figure}[!t]
%	\centering
%	\includegraphics[width=2.5in]{myfigure}
%	\input{figures/myfigure.tex}
%	\caption{text}
%	\label{fig:figure_example}
%\end{figure}

\section{DDPG Approach}\label{sec:deep}
\begin{figure}[t]
	\centering
	\begin{tikzpicture}[scale=1.0]

\tikzstyle{A0} = [-, >={stealth}, rounded corners]
\tikzstyle{A1} = [->, >={stealth}, rounded corners]
\tikzstyle{A2} = [<->, >={stealth}, rounded corners]

\tikzstyle{A0training} = [A0, color=ccolor2]
\tikzstyle{A1training} = [A1, color=ccolor2]
\tikzstyle{A2training} = [A2, color=ccolor2]

\tikzstyle{A0evaluation} = [A0, thick]
\tikzstyle{A1evaluation} = [A1, very thick]
\tikzstyle{A2evaluation} = [A2, thick]

\tikzstyle{box} = [draw, rounded corners, align=center]
\tikzstyle{boxtraining} = [box, color=ccolor2, minimum height=1.8em]
\tikzstyle{boxevaluation} = [box, thick, minimum width=6em, minimum height=3em]

\newcommand{\xshift}{6.6}
\newcommand{\yshift}{3.0}

\node (actor)
	at (0.0, 0.0)
	[boxevaluation]
	{Actor \( \actornetworkvec \)};
	
\node (exploration)
	at (\xshift, 0.0)
	[boxevaluation]
	{Exploration};
	
\node (postproc)
	at (\xshift, -1.0*\yshift)
	[boxevaluation]
	{Post\\Processing};
	
\node (environment)
	at (0.5*\xshift, -\yshift)
	[boxevaluation]
	{Comm.\\System};
	
\node (preproc)
	at (0.0, -1.0*\yshift)
	[boxevaluation]
	{Pre\\Processing};
	
\node (buffer)
	at (.5*\xshift, -.55*\yshift)
	[boxtraining]
	{Buffer};
	
\node (critic)
	at (.5*\xshift, -.17*\yshift)
	[boxtraining]
	{Critic \( \longrewardestimsca \)};
	
\draw[A1training]
	(exploration.south) |-
	node [right] {\color{black}\footnotesize \( \actionvecnoisy \)}
	(buffer.east);
	
\draw [A1training]
	(environment) --
	node [right] {\footnotesize \( \reward \)}
	(buffer);
	
\draw [A1training]
	(preproc.north) |-
	node [left] {\color{black}\footnotesize \( \statevec \)}
	(buffer.west);
	
\draw [A1training]
	(buffer) --
	node [align=center, right] {\footnotesize update \( \parameterscritic \)}
	(critic);
	
\draw [A1training]
	(critic) --
	node [align=center, below] {\footnotesize update \\ \footnotesize \( \parametersactor \)}
	(actor);
	
\draw [A1evaluation]
	(actor) --
	node [above] {\footnotesize \( \actionvec \)}
	(exploration);

\draw [A1evaluation]
	(exploration) to (postproc);
	
\draw [A1evaluation]
	(postproc) to (environment);
	
\draw [A1evaluation]
	(environment) to (preproc);
	
\draw [A1evaluation]
	(preproc) to (actor);

\end{tikzpicture}
	\caption
	{%
		The DDPG process flow consists of two main loops: The \emph{scheduling loop}~(outside, solid lines) handles the task of generating a valid allocation action, as well as gathering the system feedback to that allocation.
		Based on the experiences collected, the \emph{learning loop}~(blue, thin lines, on the inside) first tunes a critic~\( \longrewardestimsca \) that judges the goodness of an allocation action, and then uses the critic to update the actors behaviour.
	}%
	\label{fig:ddpg_architecture}
\end{figure}
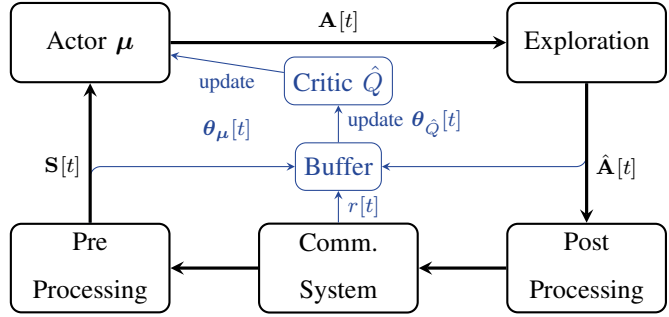

We make use of deep \reinforcementlearning in order to find an allocation algorithm that optimizes for a diverse set of goals, balancing multiple global metrics while also protecting priority users with its choice of allocation actions.
An allocation action is a vector that specifies the percentage~\( \percentageuser \) of total available resources ~\( \numresourcesavailable \) that each \usertext should receive,
\begin{align}
	\actionvec
	=	\left[
	\percentageuser[\num{1}],\quad
	\percentageuser[\num{2}],\quad
	\dots,\quad
	\percentageuser[\numusers]
	\right]
	.
\end{align}
Central to this is a \ddpg~\cite{lillicrap2015continuous} actor-critic learning module. 
It consists of a \emph{critic}, who learns to assess the quality of an allocation action according to the sum utility defined in~\refeq{eq:reward}, and an \emph{actor} that learns to output allocation actions that satisfy the critic.
Both actor and critic are modeled by standard fully connected deep neural networks, learning via variations on stochastic gradient descent from a memory buffer.
During the learning process, an exploration module introduces noise to the allocation action to ensure that the action space is properly explored and the learning process has a rich and varied set of experiences to draw on.
The learning module is bookended and assisted by pre- and post-processing blocks that embed the learning module input and output with the greater communication system, as well as ensuring valid allocation solutions.
\reffig{fig:ddpg_architecture} displays the general process flow of the scheduler for one \timesteptext, with the scheduling loop in solid lines on the outside, and the learning loop in thin blue lines on the inside.
In the following, we will describe the modules in more detail.

\subsection{The Scheduling Loop}
We start with the preprocessor that summarizes the current state of the simulated communication system in a vector~\( \statevec \) of fixed dimensionality, fit as an input to a neural network.
For each \usertext, we extract~\( \num{3} \) relevant features:
\begin{enumerate}
	\item The current channel power fading
%		\\ \( \statescauser[\num{1}] = \channelqualityuser \)
		\begin{align}
			\statescauser[\num{1}] = \channelqualityuser
		\end{align}
	\item The sum of resource requests
%		\\ \( \statescauser[\num{2}] = \sum_{\jobindex \in \jobsuserset} \resourceblockrequestedjob \)
		\begin{align}
			\statescauser[\num{2}] = \sum_{\jobindex \in \jobsuserset} \resourceblockrequestedjob
		\end{align}
	\item The inverse minimum time-to-timeout
%		\\ \( \statescauser[\num{3}] = \min_{\jobindex \in \jobsuserset} \timetotimeoutjob \)
		\begin{align}
			\statescauser[\num{3}] 	= \left( \min_{\jobindex \in \jobsuserset} \timetotimeoutjob \right)^{\num{-1}}
									= \left( \lowesttimetotimeoutuser \right)^{\num{-1}}
		\end{align}
\end{enumerate}
The resulting state vector~\( \statevec \) is saved into a memory buffer as part of an experience, for later use in the learning loop.

For a given state~\( \statevec \), the actor neural network is a function~\( {\actornetworkvec(\statevec, \parametersactor)} \) that outputs an allocation~\( \actionvec \) given its current parametrization~\( \parametersactor \).
To ensure that the output vector is normalized, \ie~\({
	\sum_{\userindex=1}^{\numusers} \percentageuser = \num{1}
}\), we use a softmax-activation for the last layer of the neural network.
Initially, the parameters~\( \parametersactor \) are randomized, and subsequently tuned to output ``good'' allocations during the learning loop.
Next, an exploration module takes the network action~\( \actionvec \) and adds random distortion~\( {\explorationnoisemultiplier \cdot \distortionuser} \) to each entry~\( \percentageuser \), with \({
	\distortionuser \sim \text{U}( \num{-0.5},\,\num{0.5} )
}\)~drawn from a uniform distribution, and a noise magnitude multiplier~\( \explorationnoisemultiplier \) that is decayed to~\( \num{0} \) over the course of training.
The resulting noisy allocation~\( \actionvecnoisy \) is clipped to the valid range of~\( [\num{0},\,\num{1}] \), re-normalized, and added to the memory buffer to for learning.

Finally, the post-processing block represents a form of incorporating model-knowledge into the learning process by, \eg limiting network output to plausible actions, filtering undesired behavior, or optimizing the output in some way that is known to be favorable.
While this can relax the burden of the learning module to learn outputs that are perfect, it also changes the learning objective from \emph{finding the best allocation solution} to \emph{finding the best post-processing input}.
An overly zealous post-processor may lead to local optimum solutions that the \deeplearning process cannot escape from.
Considering this, we opt for the following post-processing steps:
\begin{enumerate}
	\item Zero out allocations to users with no requests (implausible)
	\item Limit allocation to at most the amount requested per user (implausible)
	\item Distribute at most one additional discrete resource to any user that had a fractional resource allocated, removing the need to learn perfect discrete mappings
\end{enumerate}
During evaluation, when the learning module parameters are frozen and the scheduler is being tested for performance, we enable a further post-processing step:
\begin{enumerate}[resume]
	\item Distribute remaining available resources by order of largest remaining request, to ensure no resource remains unused due to, \eg rounding.
\end{enumerate}
After processing, the action is forwarded to the communication system, where the available resources are distributed accordingly to jobs from the queue.
One-by-one, requests from jobs are scheduled, starting with the jobs closest to timing out.
The success in the metrics defined in \refsec{sec:problemstatement} is calculated, and the simulation state updates to the next discrete \timesteptext.
The sum utility achieved by an allocation, according to~\refeq{eq:reward}, is also added to the memory buffer.

\subsection{The Learning Loop}

During each full scheduling loop iteration, the scheduler collects an experience tuple
\begin{align}
	\experience
		= 	\left(
				\statevec, \quad
				\actionvec, \quad 
				\reward
			\right)
		,
\end{align}
thereby building a data set of state-action combinations and their resulting benefit~\( \reward \).
The \ddpg algorithm uses this data set to teach a critic neural network~\( \longrewardestimsca \), parametrized by~\( \parameterscritic \), to estimate the reward resulting from a state-action combination, \ie~\({
	 \longrewardestimsca ( \statevec, \actionvec, \parameterscritic ) = \rewardestimate
}\).
In each iteration of the learning loop, a mini-batch of~\( \batchsize \) experiences is sampled from the experience buffer, and a mean squared estimation loss
\begin{align}
	\losscritic
	= 	\sum_{\batchindex = 1}^{\batchsize}
			(\reward[\batchindex] - \rewardestimate[\batchindex])^{\num{2}}
\end{align}
is evaluated on the samples~\( \batchindex \) from this batch.
Sampling is done prioritizing experiences that are new or diverge most from the estimation, as described in~\cite{hessel2018rainbow}.
Using variants on mini-batch stochastic gradient descent, the critic parameters~\( \parameterscritic \) are then updated using the gradient~\( \nabla_{\parameterscritic}\losscritic \) of this batch loss, thus nudging the critic estimate~\( \rewardestimate \) closer to the true sum utility~\( \reward \).

The relevance of the critic lies in approximating the unknown system dynamics that lead from allocation action to metric success, making them tractable as the concrete parametrized function~\({ \longrewardestimsca(\statevec, \actionvec, \parameterscritic) }\).
It is used to update the actors parameters~\( \parametersactor \) by the same principle.
A loss function
\begin{align}
	\lossactor
	=	- \sum_{\batchindex = 1}^{\batchsize}
			\longrewardestimsca
			\left(
				\statevec[\batchindex],\ 
				\actornetworkvec
				\left(
					\statevec[\batchindex],\,
					\parametersactor
				\right),\ 
				\parameterscritic
			\right)
\end{align}
is evaluated on the same batch of experiences.
The loss assesses the sum utility~\( \rewardestimate \) estimated on batch experiences given the actors current parametrization~\( \parametersactor \), as well as the critics current understanding of the system dynamics, \ie~\( \parameterscritic \).
Updating the actor parameters~\( \parametersactor \) using the gradient~\( \nabla_{\parametersactor}\lossactor \) aims to change the parameters to maximize the \emph{estimated} rewards.

%% EXAMPLE TABLE
%\begin{table}[!t]
%	\renewcommand{\arraystretch}{1.3}
%	\caption{A Simple Example Table}
%	\label{tab:table_example}
%	\centering
%	\begin{tabular}{c||c}
%		\hline
%		\bfseries First & \bfseries Next\\
%		\hline\hline
%		1.0 & 2.0\\
%		\hline
%	\end{tabular}
%\end{table}

%% EXAMPLE FIGURE
%\begin{figure}[!t]
%	\centering
%	\includegraphics[width=2.5in]{myfigure}
%	\input{figures/myfigure.tex}
%	\caption{text}
%	\label{fig:figure_example}
%\end{figure}

\section{Performance Evaluation}\label{sec:performance}
\subsection{Implementation Details}

We configure the simulation environment with~\( {\numresourcesavailable = \num{16}} \) available resources and~\( {\numusers = \num{10}} \) connected users.
According to the profiles listed in \reftab{tab:user_profiles}, we assign a normal profile to five of the users, High Packet Rate and Low Latency to two users each, and the last user is assigned the priority \emergencyvehicletext-profile.
TX-SNR~\( {\signalpower / \noisepower} \) is set to~\( \SI{13}{\decibel} \) for all users, with a variance~\({ \variancechannelquality = 1 }\) for the fading channel amplitudes.
Both training and evaluation are carried out with \( \num{50} \)~simulation steps~\( \timeindex \), repeated over \( \num{10000} \) episodes.
The job generation probability per step per user is set to~\( \chancejobs = \num{0.2} \), putting an expected load of about \( \num{1.6} \) requests per resource on the system.
We set the reward weightings of~\refeq{eq:reward} such that the submetrics have approximately equal influence on the sum metric according to their expected magnitudes, \( {\weightrewardcapacity = \weightrewardpacketrate = \num{.25}} \) for sum capacity~\( \rewardcapacity \) and sum packet rate~\( \rewardpacketrate \), and~\( {\weightrewardtimeout = \weightrewardtimeoutev = \num{1}} \) for sum timeouts~\( \rewardtimeout \) and \emergencyvehicletext-timeouts~\( \rewardtimeoutev \).
As \emergencyvehicletext-timeouts are part of the global sum timeouts, this weighting means that \emergencyvehicletext-timeouts have double the weight of other timeouts.

The neural networks for actor and critic are implemented as standard, fully connected feed-forward neural networks with \(\num{6}\)~layers each, and \( {[\num{300},\, \num{300},\, \num{300},\, \num{300},\, \num{400},\, \num{300} ]} \)~nodes per layer.
Optimization is done on batches of size~\({ \batchsize = \num{128} }\) using the Adam optimizer~\cite{kingma_adam_2015} with default parameters and a learning rate of \(\num{1e-4}\) and \( \num{1e-5} \) for critic and actor, respectively.
Exploration noise is multiplied by a factor~\( {\explorationnoisemultiplier = \num{1.5}} \) initially, decaying linearly to~\(\num{0}\) after \(\SI{50}{\percent}\) of episodes.
The learning loop is enabled after a minimum of~\( \num{100} \) experiences have been collected in the buffer.

We use Python to implement the simuation, using primarily the Tensorflow library.
Simulations are run on generic hardware.
For further implementation details, we refer to the full code, available at~\cite{gracla2020code}.

\begin{table}[t]
	\renewcommand{\arraystretch}{1.3}
	\caption{User Profiles}
	\label{tab:user_profiles}
	\begin{center}
		\begin{tabular}{c|c|c}
			\hline
			& \textbf{Delay} \( \timetotimeoutjob \) & \textbf{Max Job Size} \( \resourceblockmaxuser \) \\ 
			& in sim. steps		& in res. blocks	 \\\hline\hline
			Normal				& \( \num{20} \)				& \( \num{30} \) \\
			High Packet Rate	& \( \num{20} \)				& \( \num{40} \) \\
			Low Latency			& \( \num{2} \)					& \( \num{8} \) \\
			Emergency Vehicle	& \( \num{1} \)					& \( \num{16} \) \\
			\hline
		\end{tabular}
	\end{center}
\end{table}

\subsection{Benchmark Algorithms}
Three basic model-based scheduling algorithms~\cite{schmidt2018analyse} are implemented to gauge the \deeplearning~schedulers performance on the given application.

\begin{itemize}
	\item A Max Throughput\,(\maximumthroughput) scheduler allocates as many resources as requested to users by order of channel quality.
	\item A second scheduler operates by the max-min-fair principle\,(\maxminfair), allocating at most an equal share of available resources to each requesting user.
	\item A Delay Sensitive\,(\delaysensitive) scheduler assigns a channel priority 
	\begin{align*}
		\channelpriority
		=
			\frac{
				\packetrateuser
			}
			{
				\sum_{\userindexalt = \num{1}}^{\numusers} \packetrateuser[\userindexalt]
			}
			\frac{
				\channelqualityuser
			}
			{
				\sum_{\userindexalt = \num{1}}^{\numusers} \channelqualityuser[\userindexalt]
			}
	\end{align*}
	given each users relative channel quality~\( \channelqualityuser \) and packet rate~\( \packetrateuser \) (see \refeq{eq:packetrate}).
	Using the lowest time-to-timeout~\( \lowesttimetotimeoutuser \), and the total timeouts~\( {\totaltimeouts = \sum_{\timeindexalt=1}^{\timeindex} \rewardtimeoutuser[\timeindexalt]} \) specific to user~\( \userindex \), the scheduler also assigns a timeout urgency
	\begin{align*}
		\timeoutpriority
		=
			\frac{
				\totaltimeouts
				/
				\lowesttimetotimeoutuser
			}
			{
				\sum_{\userindexalt = 1}^{\numusers}
				\totaltimeouts[\userindexalt]
				/
				\lowesttimetotimeoutuser[\userindexalt]
			}
		.
	\end{align*}
	Both priorities are weighted and normalized, after which the scheduler assigns a share of the available resources equal to the weighted combined priority vector.
	Uniquely, this scheduler is allowed to not allocate to a job if it cannot be completed before timeout.
\end{itemize}

Random scheduling by selecting a normalized action~\( \actionvec \) with entries drawn from~\( {U( \num{0}, \num{1} )} \) serves as the baseline benchmark.

\begin{figure*}[t]
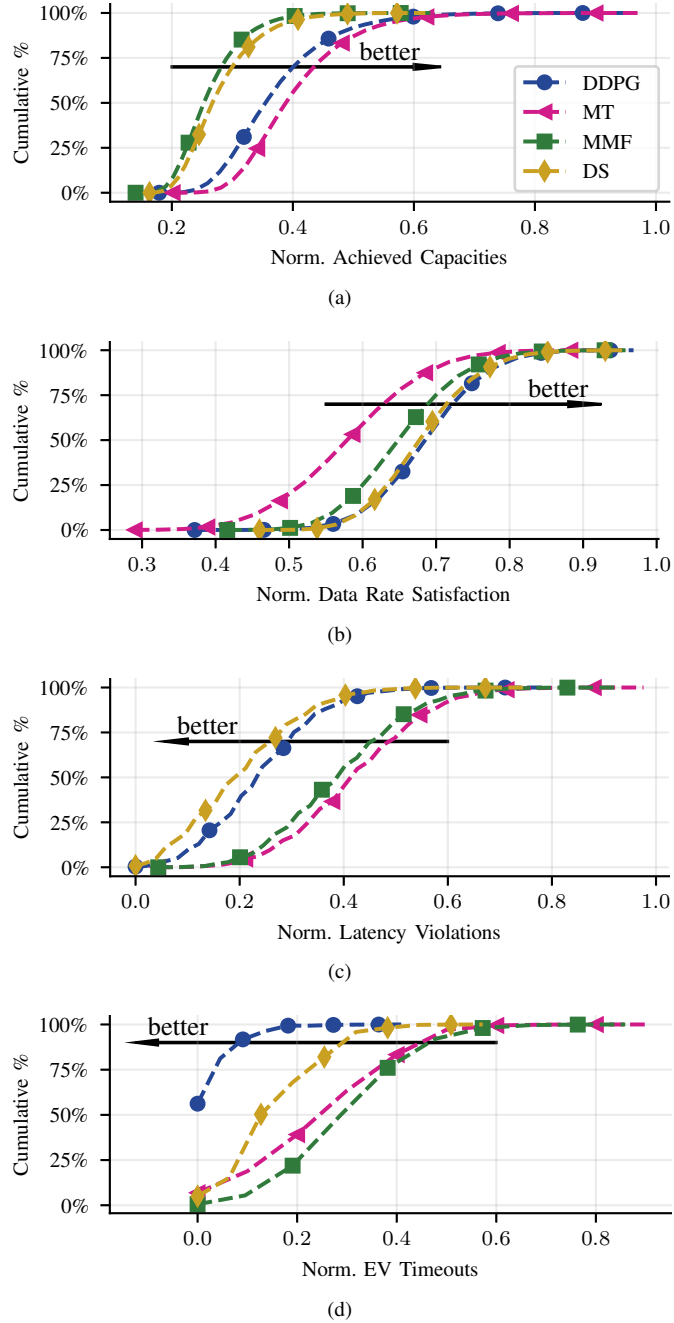

	\centering
	\subfloat[]{
		\input{figures/ddpg_scheduling_testing_capacities.tex}
		\label{fig:results_capacity}
	}
	\hfil
	\subfloat[]{
		\input{figures/ddpg_scheduling_testing_datarate_satisfaction.tex}
		\label{fig:results_packet_rate}
	}
	\\
	\subfloat[]{
		\input{figures/ddpg_scheduling_testing_latency_violations.tex}
		\label{fig:results_latency_violations}
	}
	\hfil
	\subfloat[]{
		\input{figures/ddpg_scheduling_testing_timeouts_EV_only.tex}
		\label{fig:results_ev_latency_violations}
	}
	\caption{Cumulative histograms of performance of Maximum Throughput~(MT), Max-Min-Fair~(MMF) and Delay Sensitive~(DS) schedulers as well as DDPG scheduler on submetrics of capacity~(a), packet rate~(b), timeouts~(c) and EV timeouts~(d). The DDPG scheduler achieves strong performance on all submetrics, balancing them against each other in order to maximize their sum utility.}
	\label{fig:results_submetrics}
\end{figure*}

\subsection{Numerical Results}

\begin{figure}[t]
	\centering
	\input{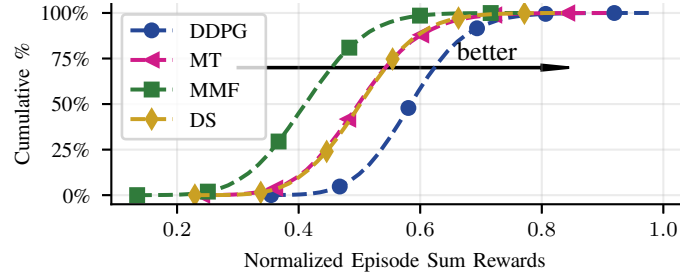}
	\caption{%
		Cumulative histogram of performance of Maximum Througput~(MT), Max-Min-Fair~(MMF), Delay Sensitive~(DS) and DDPG scheduler on the weighted sum metric.
		For each episode, all achieved rewards~\( \reward \) are summed.
		Achieved reward sums are normalized by the highest achieved reward sum.
	}%
	\label{fig:results_reward}
\end{figure}

The simulation setup inherently exhibits a variance in the performance boundaries achievable within a simulation run, e.g., in some runs random job creation leads to longer queues and subsequently more unpreventable timeouts.
Therefore, we display results in a normalized cumulative histogram of performances achieved during all simulation runs.
\reffig{fig:results_reward} shows performance on the weighted reward sum and \reffig{fig:results_capacity} to \reffig{fig:results_ev_latency_violations} break down performance on each component metric.

The \ddpg scheduler, with the help of post-processing, is able to find a scheduling approach that outperforms the simple model-based algorithms on the given reward function and weighting.
Analysis of the submetrics shows the balanced profile that the given weighting would suggest, going head-to-head with the best model-based algorithms performance in each category.
On \emergencyvehicletext timeouts specifically the \ddpg scheduler is showing by far the best performance out of any given algorithm, outperforming even the general timeout-focused Delay Sensitive scheduler, achieved by the targeted addition of \emergencyvehicletext timeouts to the maximization goal.

\begin{figure}[t]
	\centering
	\input{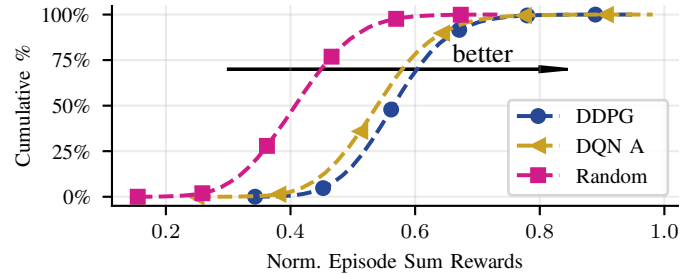}
	\caption{%
		Cumulative histogram of performance of DDPG, DQN adaptive and fully random scheduler on the weighted sum metric.
	}
	\label{fig:results_reward_adaptive_random}
\end{figure}

\reffig{fig:results_reward_adaptive_random} further compares the weighted sum metric against a random baseline and a \dqn~adaptive scheduler~\cite{gracla2021dqn} that selects from a variety of models in order to maximize a flexible reward goal.
While the \dqn ensemble scheduler is able to lean on model-based design more easily without yielding the flexible goal setting that \reinforcementlearning offers, it is ultimately constrained by its available ensemble of models.
The \ddpg~scheduler on the other hand is capable of finding a scheduling solution better than any combination of model-based algorithms, if it exists, and therefore has a higher performance ceiling.
Indeed, in the given simulation setup, the \ddpg~scheduler is able to outperform the \dqn~ensemble scheduler.

%% EXAMPLE TABLE
%\begin{table}[!t]
%	\renewcommand{\arraystretch}{1.3}
%	\caption{A Simple Example Table}
%	\label{tab:table_example}
%	\centering
%	\begin{tabular}{c||c}
%		\hline
%		\bfseries First & \bfseries Next\\
%		\hline\hline
%		1.0 & 2.0\\
%		\hline
%	\end{tabular}
%\end{table}

%% EXAMPLE FIGURE
%\begin{figure}[!t]
%	\centering
%	\includegraphics[width=2.5in]{myfigure}
%	\input{figures/myfigure.pgf}
%	\caption{text}
%	\label{fig:figure_example}
%\end{figure}

\section{Conclusion}\label{sec:conclusion}

In this work, the \ddpg~principle of \reinforcementlearning is used to learn a resource scheduling algorithm for communication system applications.
Continuous-valued output allows \ddpg to easily fit in with existing frameworks.
The scheduling strategy learned by \ddpg approximately maximizes a complex target utility, where special design boundary conditions, such as the protection of \emergencyvehicletext~traffic, can be easily incorporated without minor impact on the overall performance.
While neural-network-based learning algorithms are often regarded as black boxes, avenues for incorporating model-based expert inductive bias still exist.
Most prominently, we make use of reward design and post-processing to guide and stabilize the learning process.

\bibliographystyle{ref/IEEEtran}
{%
	\makeatletter  % prevents widows, orphans
	\clubpenalty10000  
	\@clubpenalty \clubpenalty
	\widowpenalty10000
	\makeatother
	
	\bibliography{ref/IEEEabrv,ref/references}
}%

\end{document}